\documentclass[%
reprint,
 amsmath,amssymb,
 aps,
 prx,
superscriptaddress,
]{revtex4-1}

\usepackage{indentfirst}  
\usepackage{bm}    
\usepackage{graphicx}  
\usepackage{latexsym}
\usepackage{graphicx}
\usepackage{psfrag}
\usepackage{epsfig}
\usepackage{amsmath}
\usepackage{tocloft}
\usepackage{array}
\usepackage{setspace}
\usepackage{amssymb}
\usepackage{float}
\usepackage{multirow}
\usepackage{slashbox}
\usepackage{color}
\usepackage{hyperref}
\usepackage[config=altsf]{subfig}
\usepackage{booktabs}
\usepackage{fullpage}
\usepackage[table]{xcolor}

\begin{document}

\preprint{APS}

\title{Distinctive electromagnetic signals caused by gravitational waves (of sub-solar mass primordial black hole binary mergers) interacting with galactic magnetic fields}


\author{Hao Wen}
\email[]{wenhao@cqu.edu.cn}
\affiliation{Physics Department, Chongqing University, Chongqing 401331, China.}

%
%
%
%

\date{\today}

\begin{abstract}
\indent As a candidate of dark matter, and related to many fundamental physics issues, the primordial black hole (PBH) is a crucial topic. However, so far the existence of PBHs is still not confirmed, and currently running gravitational wave (GW) detectors are still not able to distinguish them from the normal astrophysical black holes. In this article, we propose that the GWs (of PBH binary mergers) could interact with the very widespread background galactic magnetic fields  in the Milky way, to produce the perturbed electromagnetic waves (EMWs) with unique characteristics of frequencies, waveforms, spectra and polarizations. In order to be distinguished from astrophysical black holes, only the PBHs with masses less than the solarmass are considered here, and their binary mergers will radiate GWs in frequencies much higher above the plasma frequency of interstellar medium (ISM), so corresponding perturbed EMWs (in the same frequencies to such GWs) can propagate through the ISM until the Earth. 
Our estimations show that, for the sub-solar mass PBH binary mergers within the Milky way (disk or halo), the strengths of the perturbed EMWs turn into constant levels around $\sim10^{-12}$Tesla (for magnetic components) and $\sim10^{-10}Watt \cdot m^{-2}$ (for energy flux densities) at the Earth, generally for all cases of different PBH masses (and not dependent on the distance of sources), and the same mass ratio of the PBH binary gives the same strength (at the Earth) of perturbed EMWs despite different PBH masses (GW frequencies) or binary distances.
Differently, for the sub-solar mass PBH binary mergers outside the Milky way, the perturbed EMWs at Earth have lower strengths (and depend on the distance of sources), but for some part of distance range, they would also be detectable.
If such EM signals and special EM counterpart of GWs from  PBHs could be detected by space- or land-based EMWs detectors, it may provide  direct evidence of the  PBHs.
 
\begin{description}
\item[PACS numbers]
04.30.-w, 04.50.-h, 04.80.Nn, 04.30.Db
\item[Keywords] primordial black holes, gravitational waves, galactic magnetic fields, electromagnetic response to gravitational waves
\end{description}
\end{abstract}

                              
\maketitle

\section{Introduction}
\label{Introduction}
\begin{figure*}[htp]
	\centerline{\includegraphics[width=5.5 in]{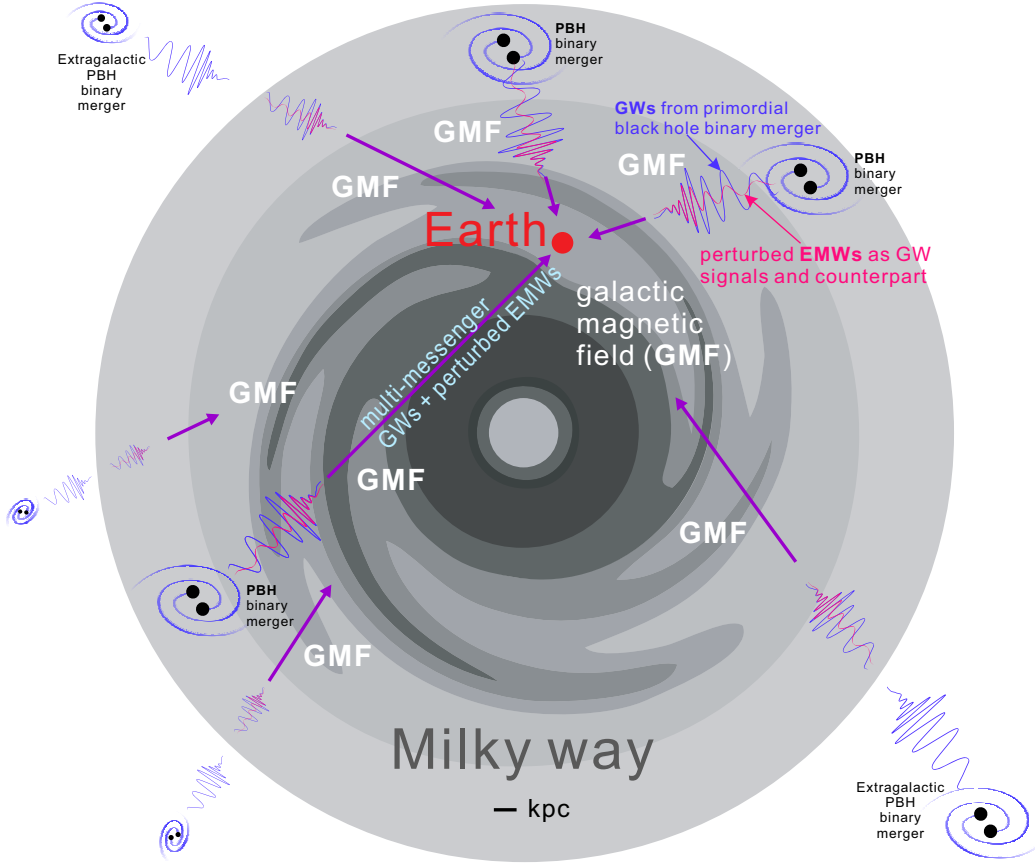}}
	\begin{spacing}{1.2}
		\caption{\footnotesize{\textbf{An intuitive view (not in scale):}
 The GWs from sub-solar mass primordial black hole (PBH) binary mergers (within or outside the Milky way) can interact with the galactic magnetic fields (GMFs, typically $\sim 10^{-10}$Tesla, widespread in Milky way), and then lead to perturbed EMWs [in the same frequencies to the GWs, e.g., $f_{\text{ISCO}}$ from $\sim$10kHz (by $\sim 10^{-1}M_{\odot}$ PBHs) to $\sim10^{20}$Hz (by $\sim 10^{-17}M_{\odot}$ PBHs); these frequencies are all far above the plasma frequency of interstellar medium (ISM) and thus the perturbed EMWs can propagate through the ISM]. The perturbed EMWs propagate until the Earth almost synchronously with these GWs, and they could be a new type of signals and special EM counterparts of GWs from the PBH binary mergers. Such EM signals   would have characteristic frequencies, waveforms (related to waveforms of the GWs, after   modifications by the ISM), spectra and polarizations. These distinctive perturbed EMWs cannot be caused by GWs from normal astrophysical black holes, and thus if they could be captured by  high sensitive land- or space-based EMW detectors, it may provide direct evidence of the PBHs.
}}
	\end{spacing}
\end{figure*}

\indent The PBHs (produced in the early Universe by collapse of large energy density fluctuations) are considered as a promising candidate of the dark matter and also related to many crucial cosmological issues and fundamental physical problems, so they have been massively studied\cite{10.1093/mnras/168.2.399,PhysRevD.81.104019,PhysRevD.94.083504,PhysRevLett.117.061101,Postnov_2019,PhysRevD.99.063523,PhysRevLett.120.191102,Raidal_2019,PhysRevD.98.123016,Chen_2018,PhysRevD.96.123523,PhysRevLett.118.151105,PhysRevLett.117.201102,Li2017,10.1093/pasj/psw065,PhysRevD.72.082002,PhysRevLett.91.021101,PhysRevD.99.103531,PhysRevD.98.043538,PhysRevLett.121.081306}. 
However, the existence of PBH is still not confirmed.
In recent years, the LIGO scientific collaboration and Virgo collaboration have reported lots of gravitational wave (GW) events from black hole mergers \cite{PhysRevLett.116.061102,secondLIGOGW,PhysRevLett.118.221101,GW170608,GW170814,LIGOnew1,PRX041015} (with frequencies around 30Hz to 450Hz and dimensionless amplitudes $\sim10^{-21}$ to $\sim10^{-22}$ near the Earth).
These black hole binaries could be formed by astrophysical black holes or primordial black holes, but so far, they  cannot  be distinguished   from these different origins.\\ 



\indent Here, we particularly focus on the PBHs with masses less than the solar mass, due to such small black holes can only be the PBHs but not astrophysical black holes.
The binary mergers of these sub-solar mass PBHs will produced GWs in higher frequencies, e.g.,  the frequency at innermost stable circular orbit  ($f_{\text{ISCO}}$) from $\sim$10kHz (by $\sim 10^{-1}M_{\odot}$ PBHs) to $\sim10^{20}$Hz (by $\sim 10^{-17}M_{\odot}$ PBHs).
If the PBHs could be one component of the dark matter, they should distribute in the vast range of Universe, within or outside the Milky way.
On the other hand, according to contemporary observations, there are very widespread background galactic magnetic fields (GMFs, strength around $\sim10^{-10}$ to $10^{-9}$ Tesla)\cite{galacticB.RevModPhys.74.775} within the Milky way.
Therefore, according to the electrodynamics in curved spacetime, in the frame of EM response to GWs, we propose that such high-frequency GWs of sub-solar mass PBH binary mergers will interact with the GMFs in the Milky way, to produce the perturbed EM waves (EMWs, in the same frequencies to such GWs), as a new type of signals and special EM counterparts of the GWs from PBHs (see Fig. \ref{Introduction}).\\

\indent The effect of EM response to GWs had been long studied\cite{Braginsky.Grishchuk.1973,Boccaletti_NuovoCim70_1970,prd2915,Chen1994,LONG1994382,FYLi.PRD2000.044018,FYLi_PRD67_2003,FYLi_EPJC_2008,FYLi_PRD80_2009,Li.Fang-Yu.120402,Li2011,PRD104025,WenEPJC2014,LiNPB2016,wenCPC2017,PhysRevD.98.064028,Li.Wen.arXiv1712.00766,PhysRevD.94.024048}, and previous works\cite{Boccaletti_NuovoCim70_1970,FYLi_PRD67_2003,FYLi_EPJC_2008,FYLi_PRD80_2009,Li.Wen.arXiv1712.00766} indicated that the strengths of perturbed EMWs depends on both strengths and spatial scales (accumulation distance) of the background magnetic fields. Thus, the GMFs will provide a huge accumulation distance to compensate the weakness of their very low strengths, and then would lead to considerable EM signals.
Besides, for the effect of EM response to GWs, the strengths of the perturbed EMWs are proportional to both the amplitude and the frequencies of the GWs. Therefore, although the amplitude of GWs from sub-solar mass PBH binary mergers are much lower than that from astrophysical black hole binary mergers, their very high frequencies  effectively compensates the weakness of their low amplitudes.\\
\indent Also, the frequencies of such perturbed EMWs are all far above the plasma frequency of the interstellar medium (ISM). Thus, they can propagate through the ISM together with the GWs until the Earth, and they would have waveforms similar to (or related to, after some modifications by the ISM) the waveforms of the GWs of PBH binary mergers. The perturbed EMWs could also have particular polarizations which may reflect some new information about tensorial and possible nontensorial polarizations of the GWs from the PBH binary mergers. If the perturbed EMWs are captured by current land-based or space-based high sensitive EMW detectors, such characteristics would be helpful to extract these EM signals from noise by methods similar to the way for searching GW signals from data of LIGO, Virgo, KAGRA, etc, of using the matched filtering based on waveform templates. Such  perturbed EMWs with unique properties cannot be caused by GWs from astrophysical black holes, so they would be direct signals of the PBHs.


\indent In Sect.\ref{estimation},  we estimate strengths at the Earth of the perturbed EMWs caused by GWs (from sub-solar mass PBH binary mergers) interacting with the GMFs, and in Sect.\ref{conclusion},  we give a short conclusion and discussion. \\

\begin{figure}[!htbp]
	\centerline{\includegraphics[width=3.5in]{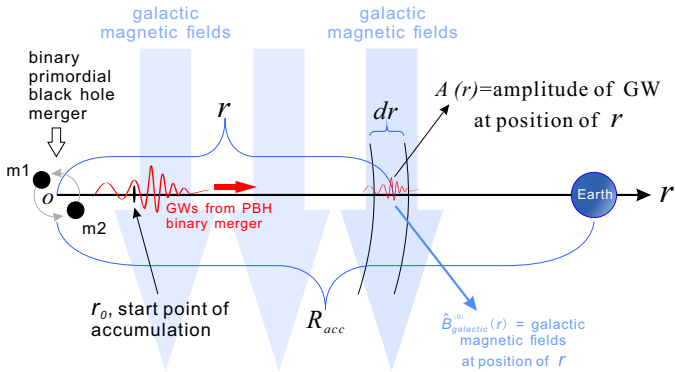}}
	\begin{spacing}{1.2}
		\caption{\footnotesize{\textbf{Calculation for accumulated perturbed EMWs caused by GWs (of sub-solar mass PBH binary mergers) interacting with GMFs.}
		}}
		\label{int}
	\end{spacing}
\end{figure}

\begin{figure*}[htp]
	\centering
	\subfigure[]{%
		\label{1a}%
		\includegraphics[width=3.1 in]{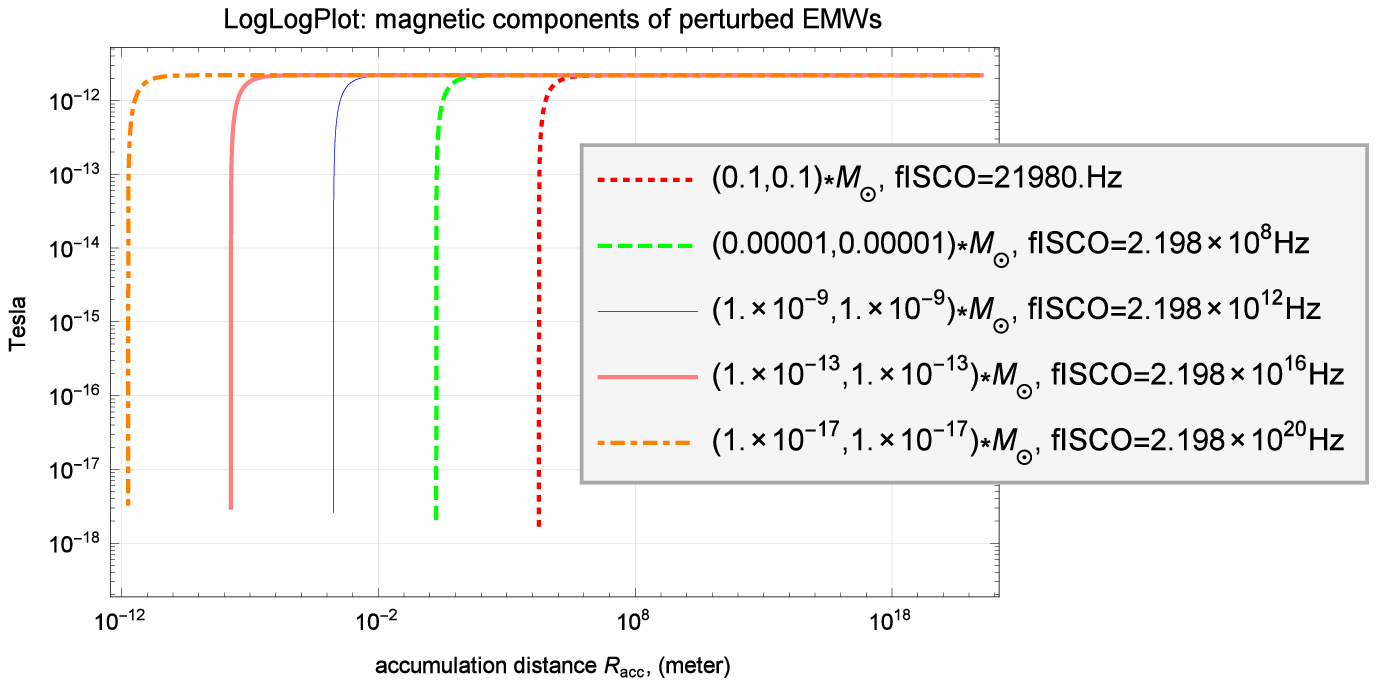}}
	\quad
	\subfigure[]{%
		\label{1b}%
		\includegraphics[width=3.20 in]{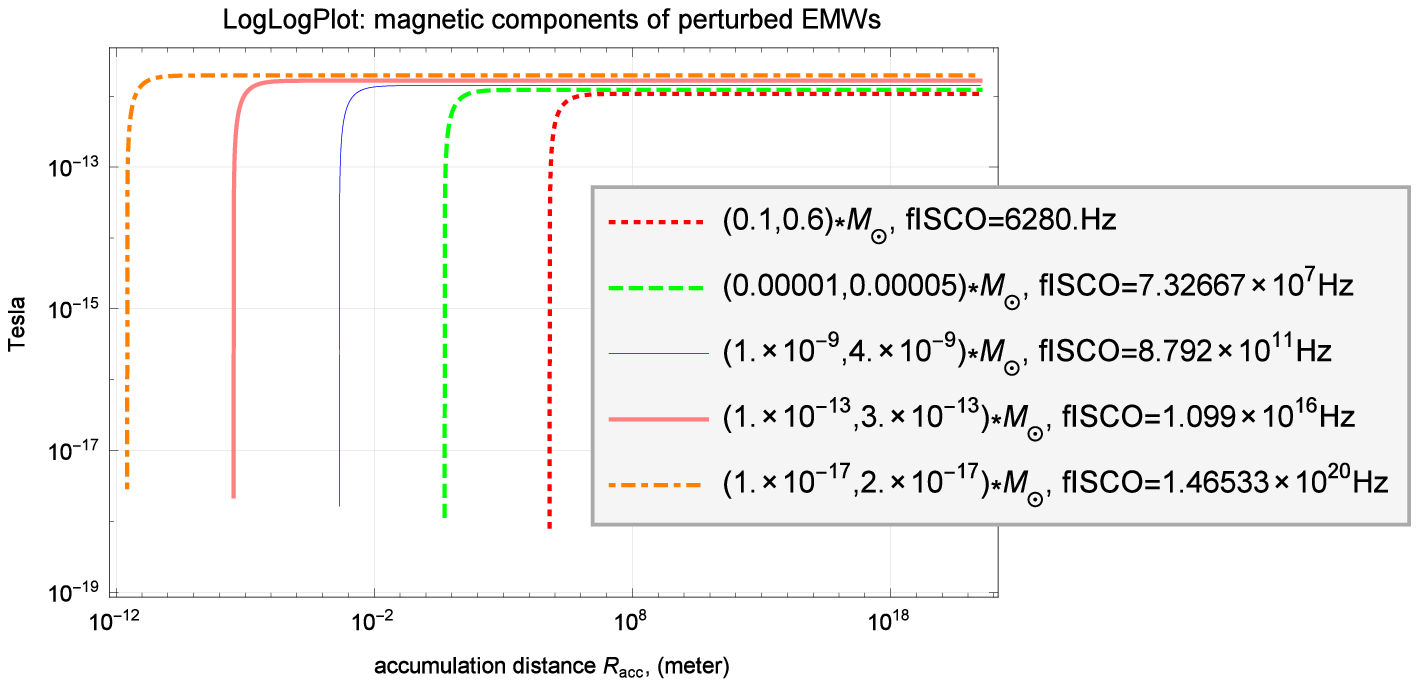}}\\
	\subfigure[]{%
		\label{1c}%
		\includegraphics[width=3.1 in]{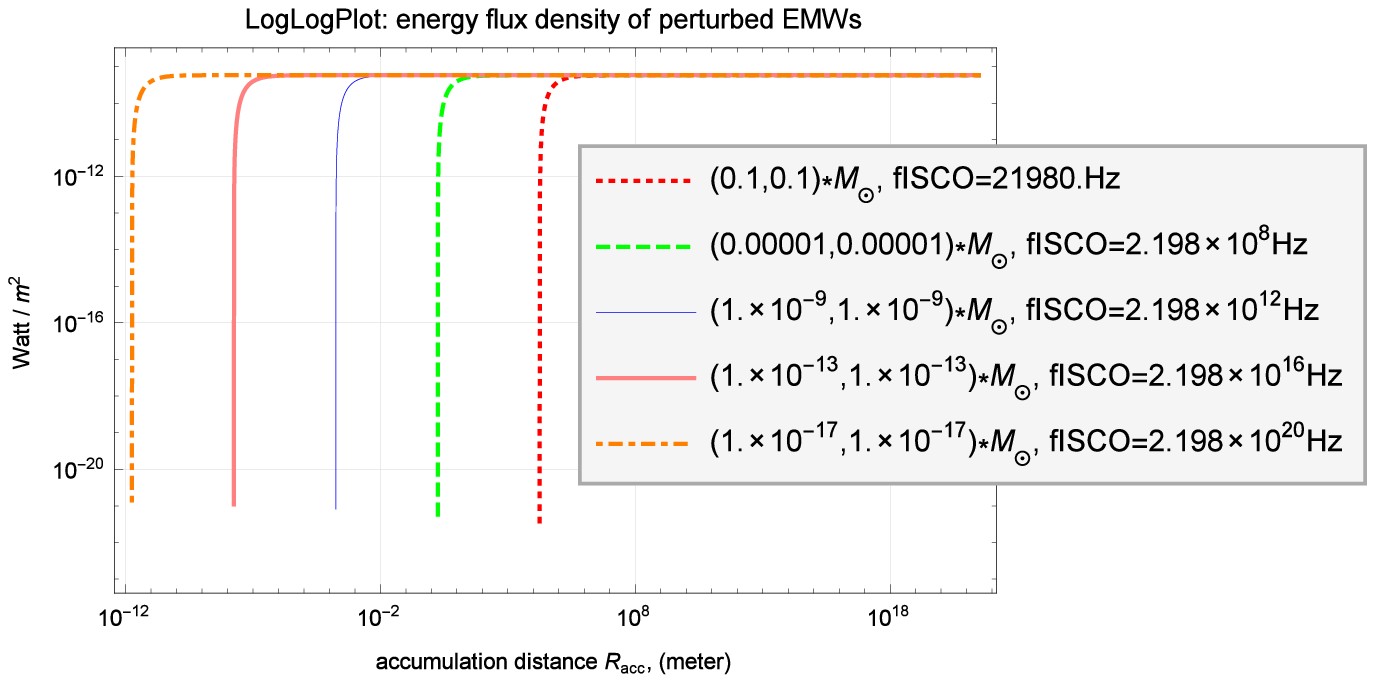}}
	\quad
	\subfigure[]{%
		\label{1d}%
		\includegraphics[width=3.20 in]{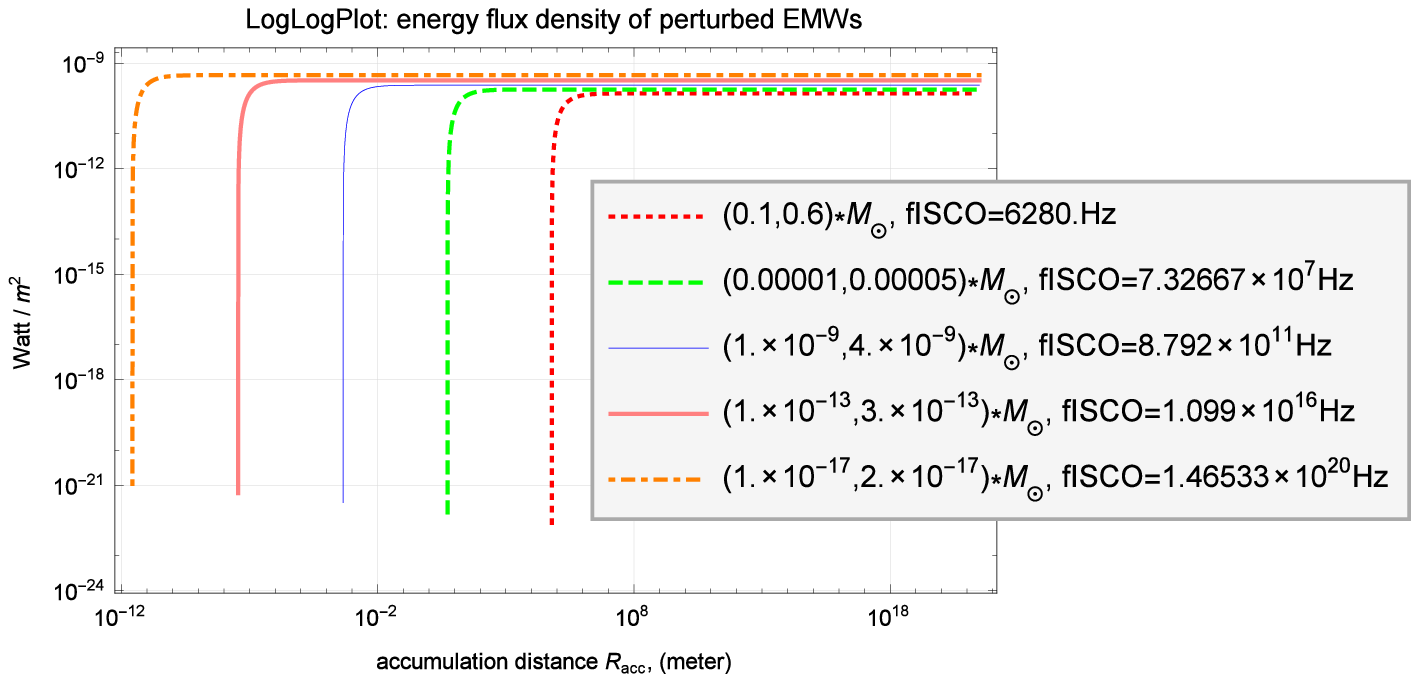}}\\
	\caption{Examples of strengths of magnetic components and energy flux densities of the accumulated perturbed EMWs caused by GWs (from sub-solar PBH binary mergers) interacting with galactic magnetic fields, for different cases of PBH masses from $\sim10^{-17}M_{\odot}$ to $\sim10^{-1}M_{\odot}$, corresponding to the $f_{\text{ISCO}}$ from $\sim10^{20}$Hz to $\sim10$kHz.  It is found the accumulated strengths turn into constant levels around $10^{-12}$Tesla (for magnetic components) and $10^{-10}Watt \cdot m^{-2}$ (for energy flux densities) until the Earth generally for all cases of different PBH masses and frequencies of GWs. The $R_{acc}$ is from the start point of accumulation $r_0$ until the Earth.}%
	\label{PLOT5accumBingalaxyB}%
\end{figure*}

\begin{figure*}%
	\centering
	\subfigure[]{%
		\label{2a}%
		\includegraphics[width=3.2 in]{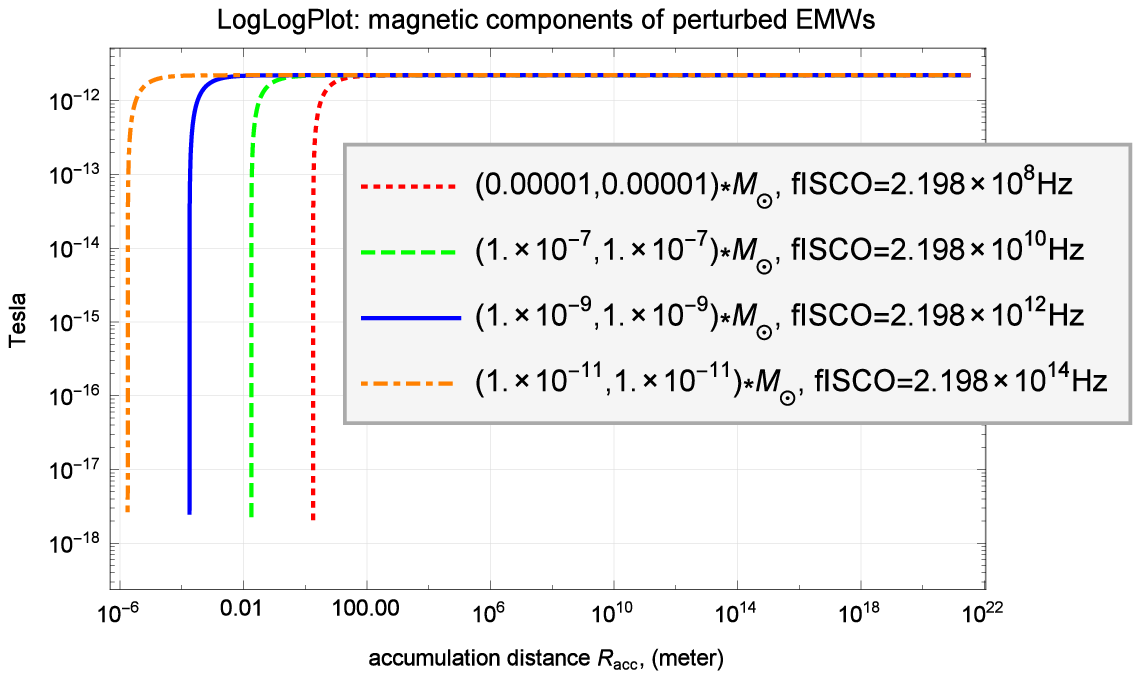}}
	\quad
	\subfigure[]{%
		\label{2b}%
		\includegraphics[width=3.05 in]{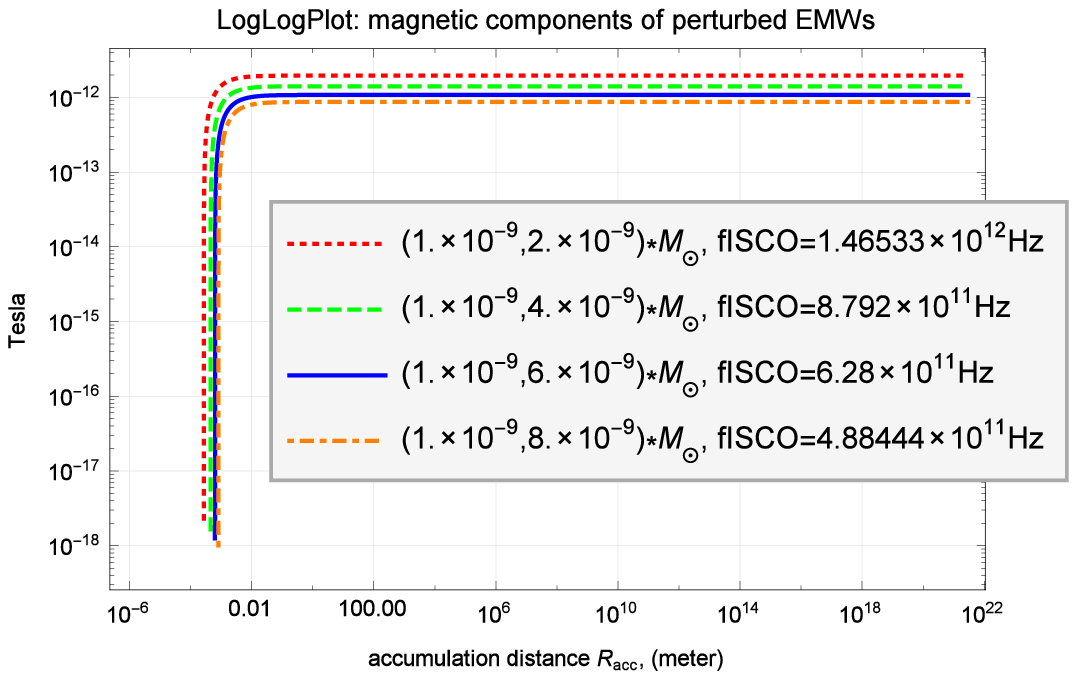}}\\
	\subfigure[]{%
		\label{2c}%
		\includegraphics[width=3.05 in]{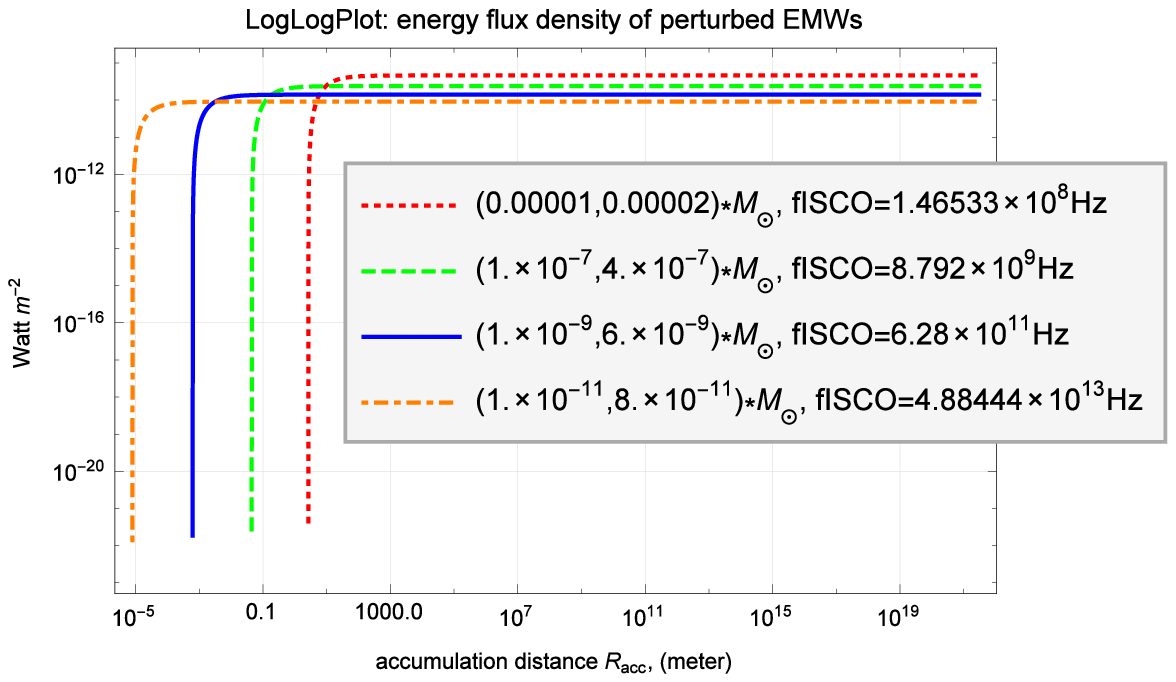}}
	\quad
	\subfigure[]{%
		\label{2d}%
		\includegraphics[width=3.25 in]{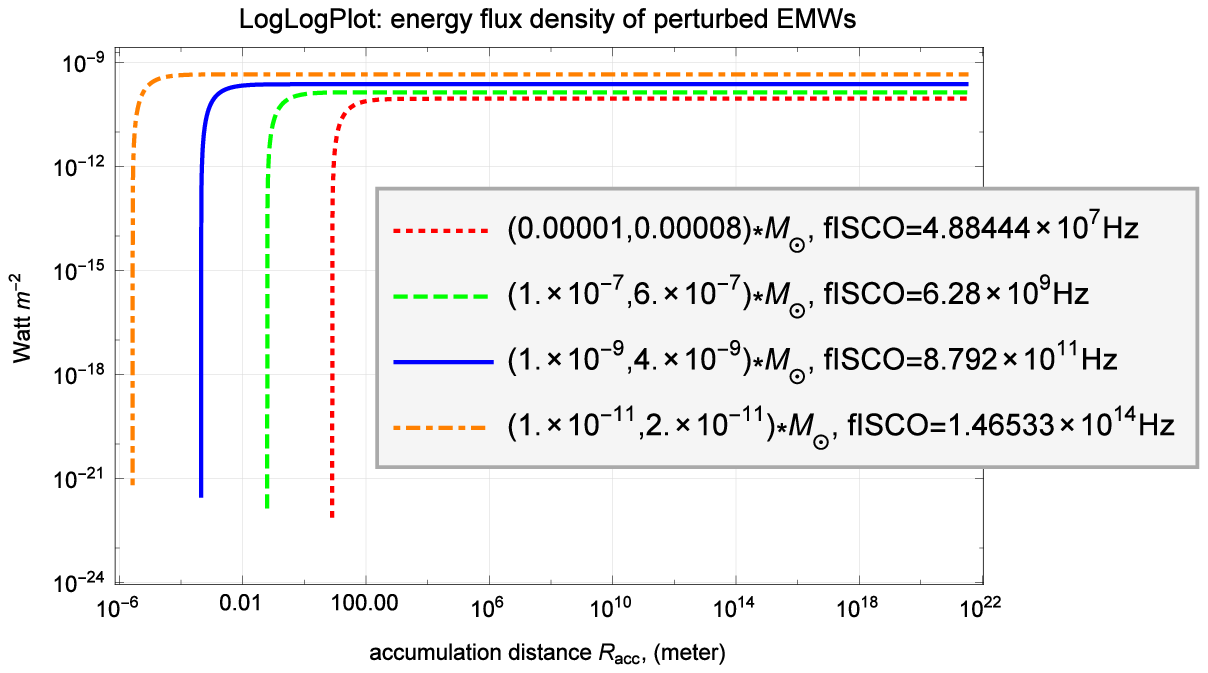}}\\
	\caption{Examples of strengths of magnetic components and energy flux densities of the accumulated perturbed EMWs caused by GWs (from sub-solar PBH binary mergers) interacting with galactic magnetic fields, for more cases of various PBH masses. It is found the accumulated strengths also generally turn into levels around $10^{-12}$Tesla (for magnetic components) and  $10^{-10}Watt \cdot m^{-2}$  (for energy flux densities) until the Earth, consistent to the Fig. \ref{PLOT5accumBingalaxyB}.}%
	\label{PLOTaccumBingalaxyB}%
\end{figure*}


\begin{figure}[!htbp]
	\centerline{\includegraphics[width=3 in]{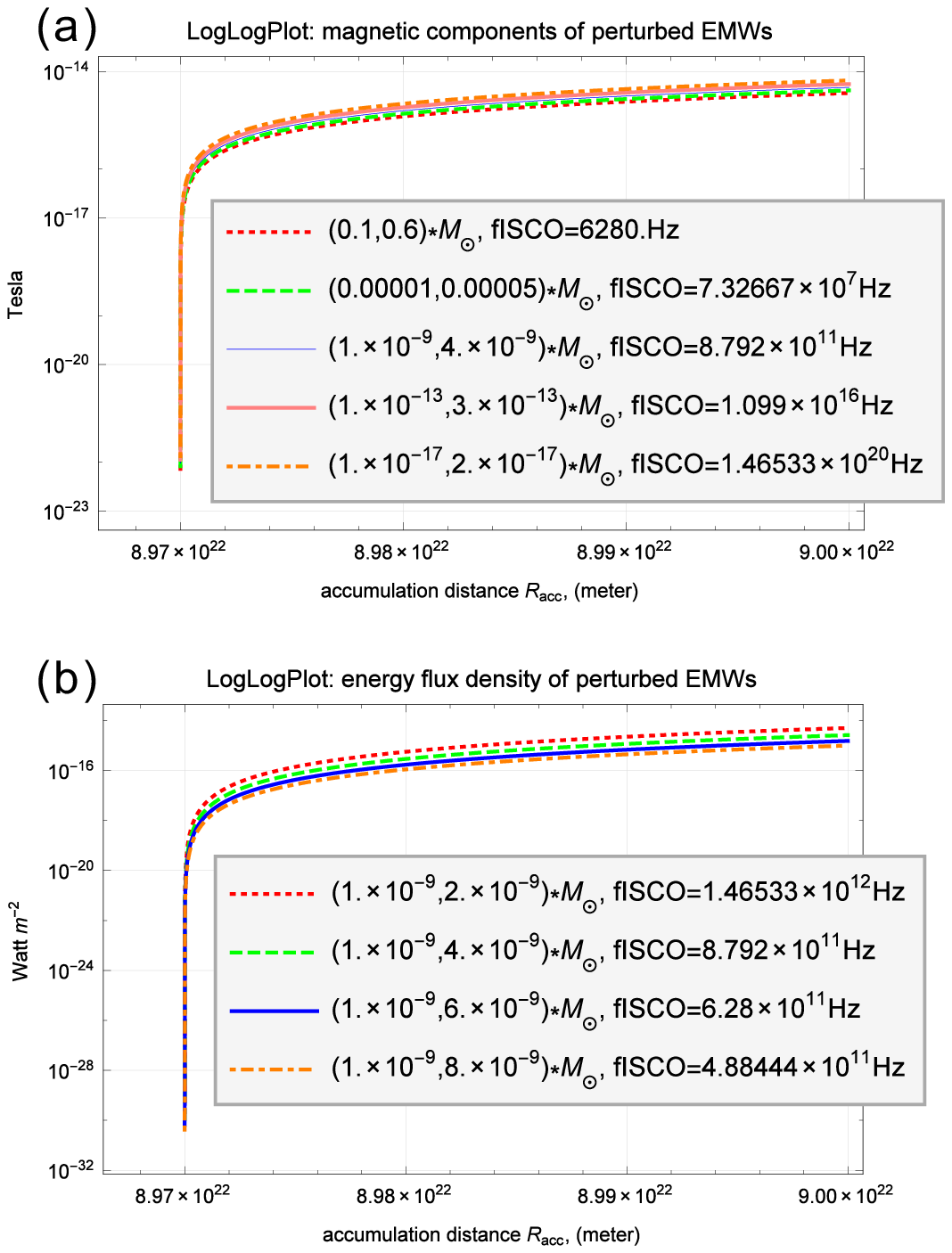}}
	\begin{spacing}{1.2}
		\caption{\footnotesize{\textbf{Similar to Figs. \ref{PLOT5accumBingalaxyB} and \ref{PLOTaccumBingalaxyB}, but for sub-solar PBH binary mergers outside the Milky way. Here, all cases have the same distance of binary mergers of 3 Mpc from the Earth.}
		}}
		\label{outsideGalaxy}
	\end{spacing}
\end{figure}

\section{Estimation of strengths of perturbed EMWs caused by GWs (of primordial black hole binary mergers) interacting with galactic magnetic fields}
\label{estimation}
Here we estimate the strengths (around the Earth) of the perturbed EMWs  caused by interaction between the GWs of sub-solar PBH binary mergers and the galactic magnetic fields in the Milky way.\\
\indent At first, we calculate the cases that the PBH binary mergers are within the Milky way (disk or halo), and later we will  extend the calculation to the cases for PBH binary mergers outside the Milky way.\\
\indent If we consider the GWs of binary PBH mergers contain not only the tensor polarizations but also possible nontensorial polarizations,  they can be generally express as:
\begin{eqnarray}\label{eq01}
&~&h_{\mu\nu} =\left( {{\begin{array}{*{20}c}
		0 &  0   & 0 &0\\
		0 & A_{+}+A_b&A_{\times}  &A_x\\
		0 & A_{\times}  &  -A_{+}+A_b& A_y\\
		0 &  A_x   & A_y& \sqrt{2}A_l \nonumber\\
		\end{array} }} \right)e^{i(\textbf{\textit{k}}_g\cdot\textbf{\textit{r}}-\omega t)},\\
\end{eqnarray}
the $+\&\times$, $x\&y$, $b\&l$ respectively represent the cross-\&plus- (tensor mode), $x$-\&$y$- (vector mode), $b$-\&$l$- (scalar mode) polarizations. Interaction of these GWs of PBH binary mergers with the GMFs in the Milky way, will generate the perturbed EMWs, and such effect can be calculated by the electrodynamics equations in curved spacetime:
\begin{eqnarray}
\label{emeqcurved}
&~& \frac{1}{\sqrt{-g}}\frac{\partial}{\partial x^{\nu}}[\sqrt{-g}g^{\mu\alpha}g^{\nu\beta}(F_{\alpha\beta}^{(0)}+\tilde F_{\alpha\beta}^{(1)})]=\mu_0J^{\mu}, \nonumber\\
\label{eq18}
&~& \nabla_{\mu}F_{\nu\alpha}+\nabla_{\nu}F_{\alpha\mu}+\nabla_{\alpha}F_{\mu\nu}=0, \nonumber\\
\label{eq19}
&~& \nabla_{\alpha} F_{\mu\nu}=F_{\mu\nu,\alpha}-\Gamma^{\sigma}_{\mu\alpha}F_{\sigma\nu}-\Gamma^{\sigma}_{\nu\alpha}F_{\mu\sigma},
\end{eqnarray}
Due to previous works\cite{Li.Wen.arXiv1712.00766,FYLi_PRD80_2009,FYLi_EPJC_2008,PRD104025,WenEPJC2014,Boccaletti_NuovoCim70_1970},  the E (electric) and B (magnetic) components of perturbed  EMWs caused by planar GWs in an interaction distance (accumulation distance) of $\Delta\textit{L}$  can be given:
\begin{eqnarray}
\label{eqEB}
\tilde{E}^{(1)}&=&A\hat{B}_{galactic}^{(0)}k_g^{}c\Delta\textit{L}\exp[{i(\textbf{\textit{k}}_g\cdot\textbf{\textit{r}}-\omega t)}],\nonumber\\
\tilde{B}^{(1)}&=&A\hat{B}_{galactic}^{(0)}k_g^{}\Delta\textit{L}\exp[{i(\textbf{\textit{k}}_g\cdot\textbf{\textit{r}}-\omega t)}],
\end{eqnarray}
here, ``$A$'' is the GW amplitude of tensorial modes ($A_{+}$, $A_{\times}$), or of nontensorial modes [here, only for ($A_x$, $A_y$), but not for ($A_b$, $A_l$), the reason is explained below]. The $\hat{B}_{galactic}^{(0)}$ can be transverse components of the galactic magnetic fields [perpendicular to direction of GW propagation, interacting with tensorial polarizations of the GWs of PBH binary mergers], or can be longitudinal components of the galactic magnetic fields [along the direction of GW propagation, interacting with vector polarizations of the GWs of PBH binary mergers\cite{Li.Wen.arXiv1712.00766}]. \\
\indent Importantly, the tensorial GWs can interact with the transverse magnetic fields but cannot with the longitudinal magnetic fields, and contrarily, the nontensorial GWs can interact with the longitudinal magnetic fields but cannot with the transverse magnetic fields.  Thus, in this article,  we only consider the vector modes of ($A_x$, $A_y$) for the nontensorial GWs, because the the longitudinal magnetic fields can only interact with ($A_x$, $A_y$) GWs and cannot interact with $A_b$ or $A_l$ GWs\cite{Li.Wen.arXiv1712.00766}.\\

\indent To  estimate the strengths, we can calculate by the scheme explained in Fig. \ref{int}, where the binary evolution is already very close to the merger time ($t=0$, defined as the time when the amplitude of GW reaches the maximum), e.g., only several periods before the separation reaching the  innermost stable circular orbit (ISCO). As shown in  Fig. \ref{int}, we can integrate the contributions [given by Eq. (\ref{eqEB}), replace the $\Delta L$ by $dr$] of generation of the perturbed  EMWs of every small accumulation distance ``$dr$'' (in such very small distance the spherical GWs can be treated as planar waves), from the  $r_0$ (start point of the accumulation, e.g., set as 10 times of ISCO)  until end point   of accumulation distance $R_{acc}$ (this ``$R_{acc}$'' is from the start point  $r_0$ to the position of Earth, or the source-Earth distance  $R_{pbh}$, and $r=0$ means the position of PBH binary). Every part of contribution of the perturbed EMWs in the ``$dr$'' will decay from the position $r$ until the end point of accumulation $R_{acc}$, so there will be a term of ``$r/R_{acc}$'' in the formula, see below.  Therefore, we work out the accumulated perturbed EMWs in the form:  
	\begin{eqnarray}
	\label{B_dipole_tensor}
	&~&\tilde{B}_{prtbd}^{(1)}=\int^{R_{acc}}_{r_0} A(r)  \hat{B}_{galactic}^{(0)}(r) \frac{\omega}{c}\frac{r}{R_{acc}} dr,  
	\end{eqnarray} 
where, the subscript ``$prtbd$'' means ``perturbed EMWs''. The  $\omega$ is the angular frequency.  The  term  $A(r)$ represents the amplitude of GWs at the position of $r$, and this can be expressed by\cite{CreightonBook,zhaowen}:
	\begin{eqnarray}
\label{hr}
&~&A(r)= 4.0\times 10^{-23}(\frac{10kpc}{r})(\frac{M_c}{0.52M_{\odot}})^{5/3}(\frac{f}{10^{-3}Hz})^{2/3},  \nonumber\\
\end{eqnarray}
where, $M=m1+m2$ (mass of two PBHs in the binary), $\mu=m1*m2/M$, $M_c=\mu^{3/5}M^{2/5}$. For a conservative estimation, we can take the frequency $f$ in the above $A(r)$ as the $f_{\text{ISCO}}$ (the GW frequency at the ISCO), and $f_{\text{ISCO}}=(6^{3/2}\pi M)^{-1}$\cite{Abadie_2010}.

Actually, during the accumulation process, the galactic magnetic fields $\hat{B}_{galactic}^{(0)}(r)$ vary along the line of sight, or, it should be the function of ``r'',  but as the first step of a estimation for the order of magnitude, we can treat the  $\hat{B}_{galactic}^{(0)}$  generally in level  of $10^{-10}$Tesla\cite{galacticB.RevModPhys.74.775}. \\

\rowcolors{6}{gray!25}{white}
\begin{table*}[!htbp]
	\caption{\label{table1}
		Strengths of perturbed EMWs caused by GWs (of sub-solar mass PBH binary mergers in Milky way) interacting with galactic magnetic fields (set as $10^{-10}$Tesla here). The ``B of  perturbed EMWs'' represents the magnetic component of the perturbed EMWs, and ``EFD'' means energy flux density. It is shown that the same mass ratio of the PBH binary gives the same strength (at the Earth) of perturbed EMWs  despite different PBH masses (GW frequencies) or binary distances.}	
	\begin{tabular}{ccccccc}
		\hline
		\hline
		PBH binary &$f_{\text{ISCO}}$ &\multicolumn{3}{c}{$A(r)$ at Earth of GWs from PBH}&B of  perturbed~~&EFD  of perturbed\\
		masses& of GWs&\multicolumn{3}{c}{ binaries with various distance}&   EMWs at  & EMWs at  \\
		\cline{3-5}
		($M_{\odot}$)&(Hz)&10 kpc&50 kpc&100 kpc&  Earth (Tesla)&  Earth ($W/m^2$)  \\
		\hline
		~\\	
		$10^{-1}$, $10^{-1}$&$2.2\times10^{4}$&$1.6\times10^{-19}$&$3.2\times10^{-20}$&$1.6\times10^{-20}$&$2.2\times10^{-12}$&$5.8\times10^{-10}$ \\
		$10^{-5}$, $10^{-5}$&$2.2\times10^{8}$&$1.6\times10^{-23}$&$3.2\times10^{-24}$&$1.6\times10^{-24}$&$2.2\times10^{-12}$&$5.8\times10^{-10}$ \\
		$10^{-9}$, $10^{-9}$&$2.2\times10^{12}$&$1.6\times10^{-27}$&$3.2\times10^{-28}$&$1.6\times10^{-28}$&$2.2\times10^{-12}$&$5.8\times10^{-10}$ \\
		$10^{-13}$, $10^{-13}$&$2.2\times10^{16}$&$1.6\times10^{-31}$&$3.2\times10^{-32}$&$1.6\times10^{-32}$&$2.2\times10^{-12}$&$5.8\times10^{-10}$ \\
		$10^{-17}$, $10^{-17}$&$2.2\times10^{20}$&$1.6\times10^{-35}$&$3.2\times10^{-36}$&$1.6\times10^{-36}$&$2.2\times10^{-12}$&$5.8\times10^{-10}$ \\
		$10^{-6}$, $2*10^{-6}$&$1.5\times10^{9}$&$2.1\times10^{-24}$&$4.2\times10^{-25}$&$2.1\times10^{-25}$&$2.0\times10^{-12}$&$4.6\times10^{-10}$ \\
		$10^{-8}$, $3*10^{-8}$&$1.1\times10^{11}$&$2.4\times10^{-26}$&$4.8\times10^{-27}$&$2.4\times10^{-27}$&$1.7\times10^{-12}$&$3.3\times10^{-10}$ \\
		$10^{-10}$, $4*10^{-10}$&$8.8\times10^{12}$&$2.6\times10^{-28}$&$5.1\times10^{-29}$&$2.6\times10^{-29}$&$1.4\times10^{-12}$&$2.4\times10^{-10}$ \\
		$10^{-12}$, $4*10^{-12}$&$8.8\times10^{14}$&$2.6\times10^{-30}$&$5.1\times10^{-31}$&$2.6\times10^{-31}$&$1.4\times10^{-12}$&$2.4\times10^{-10}$ \\
		$10^{-14}$, $3*10^{-14}$&$1.1\times10^{17}$&$2.4\times10^{-32}$&$4.8\times10^{-33}$&$2.4\times10^{-33}$&$1.7\times10^{-12}$&$3.3\times10^{-10}$ \\
		$10^{-16}$, $2*10^{-16}$&$1.5\times10^{19}$&$2.1\times10^{-34}$&$4.3\times10^{-35}$&$2.1\times10^{-35}$&$2.0\times10^{-12}$&$4.6\times10^{-10}$ \\
		\hline
	\end{tabular}
\end{table*}

\rowcolors{6}{gray!25}{white}
\begin{table*}[!htbp]
	\caption{\label{table2}
	Similar to the Table \ref{table1},	but for sub-solar PBH binary mergers outside the Milky way. Such GWs propagate and decay from the far away source until the Milky way, and then interact with the GMFs (set as $10^{-10}$Tesla) in the Milky way (the accumulation distance of GMF scale taken account to the interaction, is set to only 10 kpc for conservative estimation) to produce the perturbed EMWs. The ``B of  perturbed EMWs'' represents the magnetic component of the perturbed EMWs, and ``EFD'' means energy flux density.}
	\begin{tabular}{ccccccccc}
		\hline
		\hline
		PBH binary &$f_{\text{ISCO}}$ &\multicolumn{3}{c}{$A(r)$ at Earth of GWs from PBH}&~&\multicolumn{3}{c}{EFD of perturbed EMWs at Earth for}\\
		masses& of GWs&\multicolumn{3}{c}{ binaries with various distance}&~&\multicolumn{3}{c}{binaries with various distance ($W/m^2$)}\\
		\cline{3-5} \cline{7-9} 
		($M_{\odot}$)&(Hz)& 3 Mpc&1 Gpc& 15 Gpc&~& 3 Mpc& 1 Gpc &  15 Gpc\\
		\hline
		~\\	
		$10^{-1}$, $10^{-1}$&$2.2\times10^{4}$&$5.3\times10^{-22}$&$1.6\times10^{-24}$&$1.1\times10^{-25}$&~&$6.4\times10^{-15}$&$5.8\times10^{-20}$ &$2.6\times10^{-22}$\\
		$10^{-5}$, $10^{-5}$&$2.2\times10^{8}$&$5.3\times10^{-26}$&$3.2\times10^{-28}$&$1.6\times10^{-29}$&~&$6.4\times10^{-15}$&$5.8\times10^{-20}$ & $2.6\times10^{-22}$\\
		$10^{-9}$, $10^{-9}$&$2.2\times10^{12}$&$5.3\times10^{-30}$&$3.2\times10^{-32}$&$1.6\times10^{-33}$&~&$6.4\times10^{-15}$&$5.8\times10^{-20}$ & $2.6\times10^{-22}$\\
		$10^{-13}$, $10^{-13}$&$2.2\times10^{16}$&$5.3\times10^{-34}$&$3.2\times10^{-36}$&$1.6\times10^{-37}$&~&$6.4\times10^{-15}$&$5.8\times10^{-20}$ & $2.6\times10^{-22}$\\
		$10^{-17}$, $10^{-17}$&$2.2\times10^{20}$&$5.3\times10^{-38}$&$3.2\times10^{-40}$&$1.6\times10^{-41}$&~&$6.4\times10^{-15}$&$5.8\times10^{-20}$ & $2.6\times10^{-22}$\\
		$10^{-6}$, $2*10^{-6}$&$1.5\times10^{9}$&$7.1\times10^{-27}$&$2.1\times10^{-29}$&$1.4\times10^{-30}$&~&$5.1\times10^{-15}$&$4.6\times10^{-20}$ & $2.0\times10^{-22}$\\
		$10^{-8}$, $3*10^{-8}$&$1.1\times10^{11}$&$8.0\times10^{-29}$&$2.4\times10^{-31}$&$1.6\times10^{-32}$&~&$3.6\times10^{-15}$&$3.3\times10^{-20}$ &$1.4\times10^{-22}$ \\
		$10^{-10}$, $4*10^{-10}$&$8.8\times10^{12}$&$8.5\times10^{-31}$&$2.6\times10^{-33}$&$1.7\times10^{-34}$&~&$2.6\times10^{-15}$&$2.4\times10^{-20}$ &$1.1\times10^{-22}$ \\
		$10^{-12}$, $4*10^{-12}$&$8.8\times10^{14}$&$8.5\times10^{-33}$&$2.6\times10^{-35}$&$1.7\times10^{-36}$&~&$2.6\times10^{-15}$&$2.4\times10^{-20}$ &$1.1\times10^{-22}$ \\
		$10^{-14}$, $3*10^{-14}$&$1.1\times10^{17}$&$8.0\times10^{-35}$&$2.4\times10^{-37}$&$1.6\times10^{-38}$&~&$3.6\times10^{-15}$&$3.3\times10^{-20}$  &$1.4\times10^{-22}$\\
		$10^{-16}$, $2*10^{-16}$&$1.5\times10^{19}$&$7.1\times10^{-37}$&$2.1\times10^{-39}$&$1.4\times10^{-40}$&~&$5.1\times10^{-15}$&$4.6\times10^{-20}$ &$2.0\times10^{-22}$ \\
		\hline
	\end{tabular}
\end{table*}

\indent The estimated levels (magnetic components and energy flux densities) of the perturbed EMWs are shown in Fig. \ref{PLOT5accumBingalaxyB} for very large range of masses of the PBHs,  from  $\sim10^{-1}M_{\odot}$ to $\sim10^{-17}M_{\odot}$, corresponding to the $f_{\text{ISCO}}$ from $\sim10$kHz to $\sim10^{20}$Hz. The results are interesting that, the accumulated strengths increase and asymptotically turn into constant levels around $\sim10^{-12}$Tesla  (for magnetic components) and $\sim10^{-10}Watt \cdot m^{-2}$ (for energy flux densities) until the Earth, generally for all cases of different PBH masses (and thus different GW frequencies) or different  distance of sources.
This phenomenal is on account of two reasons: (1) the composite effect of both the accumulation of perturbed EMWs caused by  GWs and the decaying of these GWs in a spherical ratio during their propagations from the source until the Earth. (2) binary mergers with lower PBH masses produce GWs with lower amplitudes, but with higher frequencies; whereas, in the frame of EM response to GWs, the strengths of the perturbed EMWs are proportional to both the amplitude and the frequencies of the GWs, i.e., higher GW frequencies  lead to higher strengths of the perturbed EMWs, and thus the effect of lower amplitudes of GWs is offset. Fig. \ref{PLOTaccumBingalaxyB} similarly shows  strengths but in a narrower frequency range, for more cases of various PBH masses (equal or unequal), and these results are consistent to the Fig. \ref{PLOT5accumBingalaxyB}. The Table \ref{table1} gives some concrete results for different parameters, and it is shown that the same mass ratio of the PBH binary gives the same strength (at the Earth) of perturbed EMWs  despite different PBH masses (GW frequencies) or binary distances. For such strengths, these EM signals from sub-solar mass PBHs would already be detectable by current space-based or land-based EMWs detectors. \\

\indent The above calculation can also be extended for the sub-solar mass PBH binary mergers outside the Milky way. Therefore, differently to Figs. \ref{PLOT5accumBingalaxyB}, \ref{PLOTaccumBingalaxyB} and Table  \ref{table1}, the Fig. \ref{outsideGalaxy} and Table \ref{table2} present the extragalactic cases. Such GWs propagate and decay from the far away source until the Milky way, and then interact with the GMFs (set as $10^{-10}$Tesla), and the accumulation distance in the Milky way is set to only 10 kpc for conservative estimation. For such cases, the interaction range (accumulation distance) does not cover the whole range of propagation of the GWs from source to Earth, but only cover a very small part of the full distance (in the last stage of the propagation to the Earth), i.e., before the GWs arrive the GMF of Milky way, we consider that there is not interaction between GWs and background magnetic fields (therefore no generation and no accumulation of the perturbed EMWs), so in such cases, the accumulation behavior is different to that for the binary mergers inside the Milky way (where the accumulation covers the whole propagation), and thus, corresponding perturbed EMWs   will be dependent on  the distance of GW sources, as shown in the Table \ref{table2}. It is clear that for the cases of binary mergers outside the Milky way, the perturbed EMWs will be generally weaker, and they would be detectable only for some part of the distance range.\\

%
%

\section{conclusion and discussion}
\label{conclusion}
\indent 

We estimate the strengths of perturbed EMWs caused by  the GWs (of sub-solar mass primordial black hole binary mergers within or outside the Milky way) interacting with the widespread galactic magnetic fields (in the Milky way disk or halo).  Frequencies ($f_{\text{ISCO}}$) of such GWs are from $\sim$10kHz (by $\sim 10^{-1}M_{\odot}$ PBHs) to $\sim10^{20}$Hz (by $\sim 10^{-17}M_{\odot}$ PBHs), and the strength of GMF is considered as $\sim10^{-10}$Tesla.\\
\indent For cases of PBH binary mergers within Milky way, the estimated strengths of the perturbed EMWs turn into constant levels around $\sim10^{-12}$Tesla (for magnetic components) and $\sim10^{-10}Watt \cdot m^{-2}$ (for energy flux densities) until the Earth, generally for all cases of different PBH masses (and not dependent to source distance, see Figs. \ref{PLOT5accumBingalaxyB}, \ref{PLOTaccumBingalaxyB}, Table \ref{table1}). It is also found that the same mass ratio of the PBH binary gives the same strength (at the Earth) of perturbed EMWs  despite different PBH masses (GW frequencies) or binary distances  (Table \ref{table1}).\\
\indent This interesting phenomenal is on account of two reasons:
(1) the composite effect of both   accumulation of perturbed EMWs  and   decaying of GWs during their propagations. 
(2) the strengths of perturbed EMWs are proportional to both   amplitude and   frequencies of the GWs, so the higher GW frequencies  lead to higher strengths of the perturbed EMWs to compensate the effect of lower amplitudes of the GWs. \\
\indent For cases of sub-solar mass PBH binary mergers outside the Milky way, differently, the strengths of perturbed EMWs are dependent on the source distance, and they are generally weaker than the case of binary mergers within the Milky way  (Fig. \ref{outsideGalaxy} and Table \ref{table2}), and would   be detectable only for some part of the distance range.\\

\indent With above estimated strengths (especially for the cases within the Milky way), these EM signals and counterpart of GWs from sub-solar mass PBH binary mergers might be detectable by current space-based or land-based EMWs detectors for various bands, even if considering the depression of the strengths of perturbed EMWs at the Earth caused by the dispersion.
Further, although they will be recorded with massive noise, they could be distinguished by methods similar to the way for searching GW signals from data of LIGO, Virgo, KAGRA, etc, of using the matched filtering based on waveform templates. However, to build such template bank of the waveforms of the perturbed EMWs,  we need to consider the impact of some other factors, e.g., the interstellar medium (ISM) along the propagation path of the perturbed EMWs, the variation of the GMFs and the inhomogeneous distribution of electron density in the ISM. The impact of ISM brings dispersion, refraction and scattering, etc. The dispersion causes the frequency-dependent dispersive delay (thus the dedispersion will be required). Refraction of signal EMWs caused by inhomogeneities in the ISM and turbulence of electron density lead to interstellar scintillation\cite{1985ApJ.288.221C}, and the presence of GMFs in the ISM also results in the Faraday rotation. Therefore, the above effects should be included to build proper template bank of the modified waveforms (or spectra) of the perturbed EMWs, with more accurate numerical calculation based on detailed astrophysical observation data. This would be quite complicated treatments, and such issues will be investigated in our subsequent works.\\
\indent Moreover, the GWs from sub-solar mass PBH binary mergers would contain tensorial polarizations and possible nontensorial polarizations, and this leads to that the perturbed EMWs caused by such GW polarizations will be different to those caused by only the tensorial GW polarizations. On the other hand, the effect of Faraday rotations and the detailed distribution of directions and strengths of the GMFs, will also influence the polarizations of the perturbed EMWs\cite{Li.Wen.arXiv1712.00766} along the propagation path from the source to the Earth. Therefore, if the specific polarizations of the perturbed EMWs would be captured and measured, we could reversely extrapolate the possible combination of proportions of all polarizations (including nontensorial ones) of the GWs from PBH binary mergers. These studies could also be carried on soon.\\

\indent Generally, if one day we can capture and distinguish such perturbed EMWs (with unique properties of frequencies, waveforms, spectra and polarizations), due to that these signals cannot be caused by GWs from astrophysical black holes, it would provide direct evidence of the primordial black holes.\\


\bibliography{WenReferenceData20180716}

\end{document}